# FolksoDrivenCloud: an annotation and process application for social collaborative networking

Massimiliano Dal Mas
me @ maxdalmas.com

*Abstract* — **In this paper we present the FolksoDriven Cloud (FDC) built on Cloud and on Semantic technologies. Cloud computing has emerged in these recent years as the new paradigm for the provision of on-demand distributed computing resources. Semantic Web can be used for relationship between different data and descriptions of services to annotate provenance of repositories on ontologies. The FDC service is composed of a back-end which submits and monitors the documents, and a user front-end which allows users to schedule on-demand operations and to watch the progress of running processes. The impact of the proposed method is illustrated on a user since its inception.**

## I. Introduction

Nowadays we can rely on Cloud computing [1] and virtualized computer resources to process huge amount of data rapidly.

Data/information is interconnected through machine interpretable information, considered in this work as a new tag structure called "Folksodriven" [2-10]. Social networks are special case of data networks. The proposed "Folksodriven Cloud" (FDC) combines concepts and technologies that enable data modeling capturing relationships while allowing communities to define ontologies on the data.
Semantic web technologies are used to find appropriate data and services while cloud computing to aggregate, process, analyze and visualize data.

Data computations can have unplanned characteristics. Organizations without an adequate budget for investments in information technology cannot afford permanent resources to invest on computations of data. Cloud computing could be the right solution for those circumstances.

The background of related technologies are briefly described in the next section. Then, a system overview of the "Folksodriven Cloud Structure" is presented, focusing on the use of the Semantic Web for the Cloud Computing it is proposed the use of a new folksonomy tag system, called "Folksodriven", for the annotation and process of documents presenting the architecture and the prototype implementations. The impact of the proposed method is illustrated on a user since its inception. Lastly, some conclusions are inferred.

## II. Cloud Computing

We can define cloud computing [1] as a computational paradigm where:

- the services (or computational resources) are available on request of the customer/user, via a local network (as private cloud [1]) or via the Internet (as public cloud [2]);
- the architecture is scalable and flexible: it is able to handle varying amounts of load, as needed;
- once the cloud is configured properly, it manages themselves, without human

---

[1] The **private** cloud is a cloud leased and managed in-house organization that uses it.

[2] In contrast to the private cloud, **public cloud** is leased and managed outside the organization that used-plays it.

intervention (unless you want to provide new services or change existing ones).

Cloud computing can be considered as the extension of Service Oriented Architecture (SOA) [3] out to cloud-delivered resources as: storage-as-a-service, data-as-a-service, platform-as-a-service.

The problem lays on determining which services, information, and processes can be provided better on Cloud.

In conclusion, we can say that a cloud to be called as such, the advertiser should be composed of the following pillars: virtualization, automation, billing and charge-back. Must also be characterized by a high degree of elasticity and scalability both upwards and downwards (i.e. both that use of the system grows, both to decrease). There must be a user-friendly service catalog that allows the user to request a service when they need it, quickly and easily.

### III. FOLKSODRIVENCLOUD STRUCTURE

Cloud Architectures design the necessary infrastructure for the internet/intranet accessible resources/services on-demand. Those are used when it is needed, given up when they are not necessary, and making them available after the end of a job [1].

There are different types of categories of cloud services, for the FDC we consider the following:
- Data as a service (DaaS)
- Storage as a service (STaaS)
- Software as a service (SaaS)
- Platform as a service (PaaS)
- Infrastructure as a service (IaaS)

More we consider the layer "Human-As-Service" described below [11].

#### A. Human as a Service

Some services are based on a huge aggregation and extraction of information provided from a huge number of people. Every user can adopt every kind of tools and technology belonging to the

---

[3] **Service Oriented Architecture** (**SOA**) is based on components that work seamlessly with each other into stand-alone services that can be accessed separately.

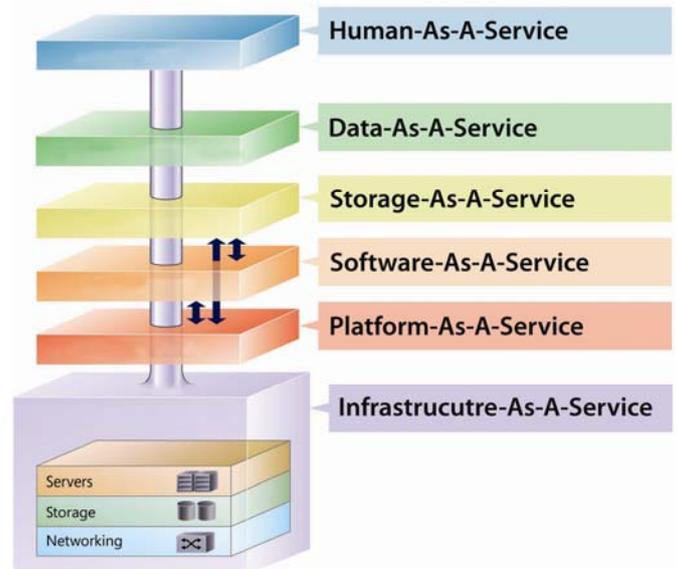

**Fig. 1** Layered architecture of cloud computing for the FDC

Crowdsourcing (CS) [4] category to achieve that task. This can be represented by a top-most layer in our stack called: Human as a Service (HaaS) [11].

HaaS is related to Human Based Computation (HBC), where a process performs its function by outsourcing certain steps to humans. These services in some cases are more targeted and controlled at promoting ideas and predicting event [12].

### IV. SEMANTIC WEB AND CLOUD COMPUTING

Semantic systems magnify data problems providing verbose data description (with RDF/OWL). A single document could result in tens of thousands of RDF/OWL triples so it is easy to reach billions of triples for a large amounts of data that needs scalability opportunities.

With unstructured data authoritative and supplementary data sources are either treated equally or require additional logic to resolve while confidence of results is reduced without addressing

Cloud provides the infrastructure needed for enterprise scalability where analytics and data are processed together to minimize network latencies using a rich analytics for expressive logic with RDF and OWL-DL.

---

[4] **Crowdsourcing** is a process where a task is outsourced to an undefined body of people (the crowd) that pursue the job list.

Cloud Semantic systems drive new User Interface (UI) capabilities as: semantic similarity analysis, dynamic faceted navigation, enterprise graph analysis, automated link analysis, social network analysis, graph matching.

Semantics in the cloud makes massive data work harder and better.

Cloud computing has emerged in recent years as new paradigm for the provision of on-demand distributed computing resources. The inherent capabilities of the cloud architecture for optimised resource management and consolidation appear very appealing for resource providers.

In this paper we present the first experiences gained by operating a community environment using the FDC cloud and the impact on users and their applications (*see Impact*).

## V. SEMANTIC ANNOTATION PROCESS

The semantic annotation process adds new metadata to the documents (in the form of RDF/OWL tags). These metadata annotations can be wider and deeper. Wide annotations do not depend on the specific subject area (e.g.: texts, figures, diagrams, …). Deep annotations are specific to one or more subject areas and are of interest to specialised searchers. We focus upon natural language annotation that can be expressed in diverse ways and are hard to find and categorize using keyword search. To achieve that we use a new forlksonomies representation on documents called Folksodriven, shown bellow.

### A. *FolksoDriven tags*

Folksonomies are trying to bypass the process of creating ontologies for semantic networks, supporting the development of evolutionary terms (and relations within a subject area) at the Human-as a-Service layer (HaaS). Folksonomies involve well-known problems and defects, such as the ambiguity of the terms and use of synonyms. Skeptics think that the folksonomy represents the first step toward anarchy of the Web but the Web is already an anarchist, which are both its greatest strength and its greatest weakness. As the Web grows, the best way to generate the metadata for the resources of the Web will always be a central issue.

Folksonomies represent exactly one of the attempts to monitor and categorize Web content to limit the costs of the "anarchy of the Internet".

The network structure of "Folksodriven tags" – Folksodriven Structure Network (FSN) – was thought as a "Folsksonomy tags suggestions" for the user on a dataset built on chosen websites.

$$(1) \quad FD := (C, R, X)$$

As stated in [2, 6] we consider a Folksodriven tag as a tuple (1) defined by finite sets composed by:

• Formal Context (C) is a triple C:=(T, D, I) where the objects T and the attributes D are sets of data and I is a relation between T and D;

• Resource (R) is represented by the uri of the webpage that the user wants to correlate to a chosen tag;

• X is defined by the relation $X = C \times R$ in a Minkowski vector space [13] delimited by the vectors C and R.

### B. *Maintaining the Integrity of the Specifications*

The primary motivation of the FDC was the belief that the semantic and the cloud technologies are complementary and their combination can offer new capabilities and optimize resource utilisation. Where cloud offers flexibility in resource management, the semantic provides high-level reasoning that enable collaboration and resource sharing among dispersed resourcers. FDC focuses on the Infrastructure-as-a-Service (IaaS) cloud paradigm, which implies the usage of virtualisation technologies for the provision of computing resources.

The project is integrating a cloud distribution, specifically designed with the purpose to host semantic services. During the design phase the specific requirements and/or restrictions of semantic services are taken into account in order to provide optimised cloud environments for deploying virtualised repositories sites.

These requirements are both technical and operational. In the heart of the FDC cloud distribution lies SOAP and WSDL.

*C. SOAP and WSDL*

WSDL represents an agreement between the service requestor and the service provider, as a Java interface represents an agreement between client code and the considered Java object. WSDL is language and platform independent and is mainly used to describe SOAP services. WSDL provides a common platform for automatically integrate different services with a common language to describe those. With WSDL, a client can locate a web service and invoke any of its available functions. While WSDL-aware tools can be used to automate processes: applications can be easily integrated on new services with no or little manual code.

## VI. FOLKSODRIVENCLOUD ARCHITECTURE

FolksoDrivenCloud (FCD) was first developed for a close experiment on an industry community of an agricultural cooperative (*see Impact*).

Prior to the turn on of that application, members of this development project community were often participating in coordinated efforts to submit a common document repository by simultaneously submitting their documents on their developing projects. Collating the results of these multi-user submissions was time-consuming, yet the site performance evaluations these submissions enabled were critical during the submitting phase of many different users. In the end, these manual document submissions motivated the development of an automated service to carry out the documents submissions and present the results via a web interface.

The FDC service is composed of a back-end which submits and monitors the documents (*see Annotation as a Service*), and a user front-end which allows users to schedule on-demand operations and to watch the progress of running processes (*see Process as a Service*). Documents are submitted and monitored using a Common Programming Interface (CPI) [5], this tool's cloud programming interface provides an efficient framework to develop a cloud service which needs the flexibility to submit arbitrary documents to any cloud back-end.

The web interface is developed using Django (https://www.djangoproject.com) and is designed to provide common web views in a core FDC library while allowing Virtual Organisations (VO) [6] to customise their web views in VO-specific plugins. Example metrics provided by FDC include job documents rates, timings of the various steps of a cloud job (e.g. preparing input files, uploading, and storing output files), and I/O metrics including storage latency and throughput values.

FolksoDrivenCloud (FDC) is an automated application service with two main goals.

First, it is a classification tool which can classify large numbers of real unstructured documents to objectively aid in the commissioning of new knowledge and to evaluate changes to knowledge structure (*see Annotations as a Service*). The service presents plots of topics update and many performance metrics to enable quick comparisons to past knowledge on a topic and between other topics (e.g. it can periodically sends short topic abstract to continually update a user on topics of interest and related cloud services).

Second, it is a functional application tool for on-demand provision of elastic computing services, semantic resource providers can optimise the utilisation of their physical computing resources (*see Process as a Service*).

In addition FDC provides a user-friendly web interface.

*A. Annotation as a Service*

We consider as service the large scale annotation that could be possible using the FDC. A general overview of the use for annotate document with the FDC among the DaaS, SaaS and HaaS is presented in Fig 2. Documents are composed by different elements (e.g.: images, tables, words, phrases) and these can be tagged by users using the folksonomy suggestion from the FD tags system.

For the aim of our prototype we have developed the necessary infrastructure for the storage, indexing, annotation and retrieval of textual and image documents. That comes from the document of the agricultural cooperative used for testing.

---

[5] **Common Programming Interface** (**CPI**) is an application programming interface (API) developed by IBM to provide a platform-independent communications interface**.**

[6] We consider a **Virtual Organization** as organization which operates primarily via electronic means sharing resources to achieve their goals.

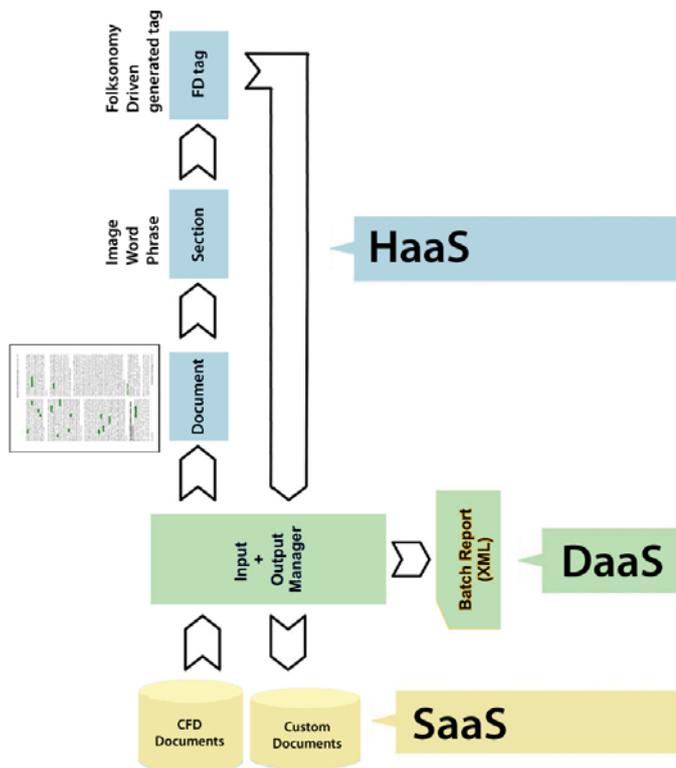

**Fig. 2**. General overview of the Annotation as a Service.

The search and retrieval functionality has been developed using the Lucene project framework (http://lucene.apache.org); this is an open-source, high-performance, full-featured text search engine library written entirely in Java. Our proposed system allows the user to search individually in selected resources or globally across a document repository using the DaaS layer, it also facilitates filtering by authors or dates by using an advanced search.

The prototype makes use of the annotation process described in papers [2-10] that make it possible for users to generate FD tags over specifisections (e.g. words, sentences) of documents. Nowadays there are over 800 documents in our system and new documents can be easily loaded. Users can share and reuse tags for improving the classification of documents. A user can also specify external resources associated with FD tags and sections of the tagged documents.

- *Prototype implementation*

The FDC prototype - for layers DaaS, SaaS and HaaS - itself allows the generation FD tags. The annotation infrastructure is based on the Monq Java software package a fast processing of text based on regular expressions implemented for Java (http://monqjfa.berlios.de). A filter server is a system that provides a service control, filtering, and caching of web traffic from users. The users' clients send a stream of text connecting to a filter server that can recognize the vocabulary of a shared terminology in the community of users. The uploaded text is recognized by an embedded Deterministic Finite Automata (DFA) [7] for automatically tagging the terminology with FD tags.

A processing pipelines can be done by a cascade of multiple filter servers - the output of one filter becomes the input of the next. The disambiguation on text is done with an acronym resolution and term frequency. The FD tags annotated information is written on an RDF/OWL file as result of the described process. The produced set of predefined FD tags can be modified anytime by users. A set of links of the updated documents is built over the defined FD tags. The defined FD tags can be automatically linked to a other sets of external resources. The FD ontologies constructed with the FD tags and the corresponding external resources, can be extended.

Users can use the clouds of FD tags to find more information and build more expressive and accurate annotations. Users can determine FD tags associated with collections of documents. According to this tag association is possible to measure how a related documents are closely. The application also allow the user to verify, validate, transform and visualize files already created to make sure they adhere to a specific format or standard.

More significant, users can contextually discover how valid the coincidence in FD tags may be by browsing through the surrounding area defined by the FD tags associated to a document.

The clouds of FD tags for each document are constantly updated by adding new FD tags and/or external resources. New external resources (as URIs or extra information) can be done by users typing those into a text field.

---

[7] A **Deterministic Finite Automaton** (**DFA**), or deterministic finite state machine, is a finite state machine that accepts/rejects finite strings of symbols and produces a unique computation for each input string.

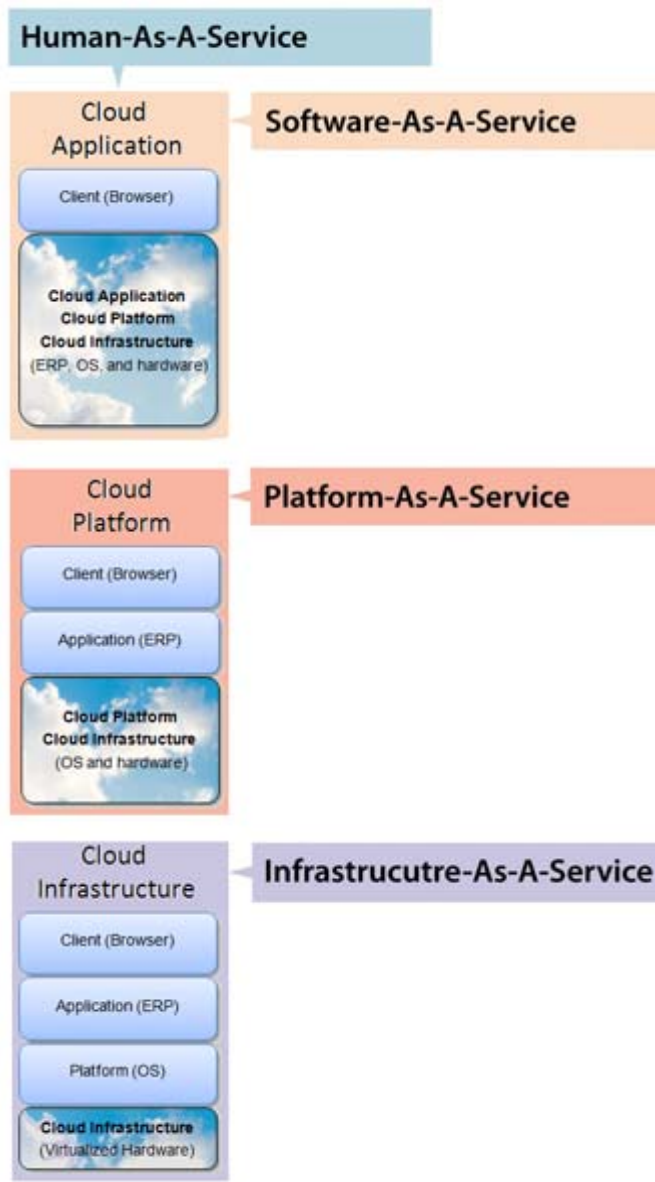

**Fig. 3**. Logical layout of the Process as a Service.

B.  *Process as a Service*

We consider the FDC provides the community at large with web based layers HaaS, SaaS and PaaS over the IaaS which allow users to schedule on-demand operations and to watch the progress of running processes (Fig.3).

Nowadays several questions often require months or even years to be solved using the computational power provided by a single cluster.

In such cases it is important to exploit the cloud in order to reduce the time execution.

Semantic technologies can significantly alter the way cloud services are currently deployed and operated. By exploiting the inherent capabilities of IaaS clouds for on-demand provision of elastic computing services, semantic resource providers – working at HaaS, SaaS and PaaS – can optimise the utilisation of their physical computing resources.

The primary user of FDC is the semantic resource administration (at SaaS), yet the adoption of the cloud paradigm is expected to indirectly have a positive impact to the VO managers and end-users.

The FDC also prepares and makes available through the appliance repository a set of VM images (at PaaS), with a semantic middleware pre-installed. The usage of VMs accelerates the instantiation of different type of plugins and makes it easier to try out new features or validate middleware updates.

The cloud computing "Process as Service" gives the ability to users to instantiate and manage VMs[8], and an appliance repository where the VM images are stored. This reference cloud service is used also internally by the developers as tester environment for deploying new plugins application and in order to investigate potential implications of their operation over the cloud.

Moreover, for VOs and end-users the ability to use VMs with pre-installed applications and software provides additional flexibility for deploying semantic services customised for specific application domains. Also, end-users are expected to be impacted positively by the expected enhanced stability and availability of plugins applications running on virtualised environments.

- *Prototype implementation*

The collaboration between users and developers has been found to be very productive: the developers takes care of splitting the application into independent tasks, filling the central database with the independent tasks, submitting the tasks to the cloud and retrieving the output. In other words, the final user does not have to deal with all the cloud technicalities which are taken by the expert.

---

[8] A **virtual machine** (**VM**) is a software implementation of a computing environment that executes programs like a physical machine. VM typically emulates a physical computing environment where a virtualization layer translates hardware and network requests to the underlying physical hardware.

We have tried to go one step behind, trying to enable final users to use the cloud on their own. For this purpose we have developed a tool, FDC, which uses a web graphic interface, written using PHP, Javascript and XSLT, where a user can authenticate using a username and password. Using the same graphic interface the user uploads the input files. A mechanism based on the https protocol has been implemented to upload input files on the Storage-as-a-Service layer in an efficient way and with an automatic registration of the files. The user can chose the plugin application depending on what he wants to do (as an analysis of the obligations of a law, or the costs and benefits for a specific farming, or an agricultural market forecasts), customise the application (changing the configuration parameters) and specify the name of the output files. After the submission of the web form, FDC executes all the steps required for the submission of the application to the cloud (slicing of the problem, user authentication by means of a certificate, submission of the jobs to the cloud, retrieving of the output files). At the end, FDC sends the link of the output files to the user by e-mail.

## VII. IMPACT

The current users of FDC include two experiments. An agricultural cooperative is the heaviest user, having used the service since its inception.

The advantage for a small community of users (as an agricultural cooperative) or a Small and Medium Enterprise (SME) is on having a public cloud on the shelf ready to be used with an affordable cost – considering that a private cloud cannot be affordable for such organizations. Meanwhile the Folksodriven classification system can help for the annotation and the research of the documents.

The agricultural cooperative represents an example of a set of users that need a collaborative environment, ready to use and flexible, not having at their internal association neither the technological know-how nor the economic power to create a "dedicated internal system".

Moreover the agricultural cooperative needed to manage documents for all the information both internal and external to the community. Internal information could be update on conferences, or consulting for the community members. External information could be: news (e.g. laws of interest), specific topics (e.g. the organic colviation trend in Italy), or calls for proposals (e.g. funding for small farmers).

The transition of FDC from a single to multiple-VO service required the generalisation of core components and the development of a plugin architecture. As a result of FDC: an automated service for storing and evaluating of cloud plugins' application work, FDC can flexibly accept further additions of plugins' applications for new communities.

The FDC service empowers administrators to undertake detailed studies of their cloud's capabilities without requiring any VO-specific knowledge or permissions. With only few clicks, FDC users can schedule a plugins' process and shortly thereafter performance metrics are made available.

The experiences of the agricultural cooperative experiment in using FDC demonstrate the potential of such a tool.

The primary focus of the agricultural cooperative using has been on optimising the data access method at the cloud. In particular, FDC was used to compare strategies such as copying input files to a local disk against reading files directly from the cloud storage element using the local access protocol. During a large scale use test, FDC was used to simulate the resource requirements of hundreds of real users by delivering a constant stream of up to 1,000 concurrent jobs throughout a two week period. Tests like this have led to I/O optimisations in the FDC software resulting in improvements to the overall job throughput on the cloud.

For functional testing, FDC delivers around ten types of test jobs for FDC. These tests validate not only the basic functionality of the cloud architecture, but are also used to test remote database access, to validate release-candidates of the cloud middleware, and to compare data-access methods. Further, a subset of the FDC functional tests are deemed as critical – consecutive failures of these jobs result in FDC taking action to blacklist the plugins from receiving user jobs. While a plugin is blacklisted,

FDC continues to send tests; when the jobs succeed again the plugin is informed and they can reset them online at their convenience.

VIII. CONCLUSION

In summary, FDC is an automated repository service which can be used for on-demand documents with continuous adjournment delivering and specific plugins for different kind of document analysis. Such a tool has been demonstrated to be useful during FD cloud creation and to evaluate changes to that.

An agricultural cooperative showed a strong interest in the service continuing to use the service for on-going documents adjurnment and their validation, while other organizations on the Small and Medium Enterprise (SME) are evaluating to use FDC operations procedures. Actually FDC has thirteen nodes that can schedule on-demand document analysis

The flexible architecture of FDC will allow future development of new plugins for different kind of communities who have interest in the service.

This distribution is used to offer a reference public cloud service which is used for VOs.

As the distribution evolves new capabilities will be added enhancing the current cloud solution. In parallel the public cloud service is attracting real-life applications which will help us benchmark the technologies and validate the applicability of FDC cloud distribution for the operation of production level.

**Massimiliano Dal Mas** is an engineer working on webservices, trafficking and online advertising and is interested in knowledge engineering. In the last years he had to play a critical role at Digital Advertising business, cultivating relationships with key publisher partners with experience managing a team. Been responsible for all day to day operations with partners and consult on the best ways to monetize their properties. His interests include: user interfaces and visualization for information retrieval, automated Web interface evaluation and text analysis, empirical computational linguistics, text data mining, knowledge engineering and artificial intelligence. He received BA, MS degrees in Computer Science Engineering from the Politecnico di Milano, Italy. He won the thirteenth edition 2008 of the CEI Award for the best degree thesis with a dissertation on "Semantic technologies for industrial purposes" (Supervisor Prof. M. Colombetti). In 2012, he received the best paper award at the IEEE Computer Society Conference on Evolving and Adaptive Intelligent System (EAIS 2012) at Carlos III University of Madrid, Madrid, Spain. In 2013, he received the best paper award at the ACM Conference on Web Intelligence, Mining and Semantics (WIMS 2013) at Universidad Autónoma de Madrid, Madrid, Spain. His paper at W3C Workshop on Publishing using CSS3 & HTML5 has been recommended as Position Paper.